# Probing Hyperbolic Polaritons using Infrared Attenuated Total Reflectance Micro-spectroscopy

Thomas G. Folland[1], Tobias W. W. Maß[2], Joseph R. Matson[3], J. Ryan Nolen[3], Song Liu[4], Kenji Watanabe[5], Takashi Taniguchi[5], James H. Edgar[4], Thomas Taubner[2] and Joshua D. Caldwell[1]

[1]Department of Mechanical Engineering, Vanderbilt University, Nashville, TN 37212

[2]Institute of Physics (IA), RWTH Aachen University, 52056 Aachen, Germany

[3]Interdisciplinary Materials Science Program, Vanderbilt University, Nashville, TN 37212

[4]Department of Chemical Engineering, Kansas State University, Manhattan, KS 66506

[5]National Institute for Materials Science, Tsukuba, Japan

**Hyperbolic polariton modes are highly appealing for a broad range of applications in nanophotonics, including surfaced enhanced sensing, sub-diffractional imaging and reconfigurable metasurfaces. The existence of this type of mode depends on extreme anisotropy in the optical properties of a natural or artificial material. However, due to their high in-plane momentum (short polariton wavelength), these hyperbolic polaritons cannot be launched by light propagating in free space. Here we show that attenuated total reflectance micro-spectroscopy (ATR) using standard spectroscopic tools can launch hyperbolic polaritons in a Kretschmann-Raether configuration. We measure multiple hyperbolic and dielectric modes within the naturally hyperbolic material hexagonal boron nitride as a function of different isotopic enrichments and flake thickness. This overcomes the technical challenges of measurement approaches based on nanostructuring, or scattering scanning nearfield optical microscopy. Ultimately, our ATR approach allows us to compare the optical properties of small-scale materials prepared by different techniques systematically.**

Hyperbolic polaritons are a class of polaritonic[1, 2] waves that are supported by highly birefringent materials where the dielectric function along orthogonal axes is opposite in sign[3, 4]. For materials exhibiting a negative real part of the permittivity (dielectric function) along one or two axes in Cartesian coordinates, the material is referred to as a Type I or Type II hyperbolic material respectively. Due to the extreme anisotropy in the optical response, hyperbolic polaritons propagate through the volume of the material, analogous to plane waves in a dielectric medium, but provide exceptional wavelength compression (optical confinement) consistent with surface polaritons. Furthermore, hyperbolic polaritons can be supported at arbitrarily large wavevectors at a given frequency, but the propagation direction within the material is dictated by the dielectric function[5-8]. This results in several effects, including negative refraction[6, 8], sub-diffractional volume-confinement of electromagnetic optical near-fields[5], and a continuum of multiple branches in the polariton dispersion[7] giving rise to an exceptionally large photonic density of states. These properties can be exploited to develop compact hyperlenses[6, 8, 9], waveguides[10, 11] and surface sensors[12, 13] far below the diffraction limit.

Traditionally hyperbolic behaviour was achieved by using artificial metamaterials consisting of superlattices of alternating sub-wavelength size metal and dielectric layers[14] or metallic nanoparticles embedded within a dielectric matrix[15]. Due to the metallic components, these suffer from significant optical losses. However natural materials can also be hyperbolic,[16] with hexagonal boron nitride (hBN) providing an excellent example[2, 5, 7]. In hBN, hyperbolicity arises in the mid-infared (MIR) due to highly anisotropic polar optical phonons, which have significantly different energies in- and out- of plane. Consequently, hBN has two spectral bands where hyperbolic polaritons can be supported, of both Type I (spectral range ~740-825 cm$^{-1}$, the 'lower Reststrahlen band') and Type II (spectral range ~1360-1614cm$^{-1}$, the 'upper Reststrahlen band'). As the hyperbolic behaviour arises from the interaction of light with phonons, the scattering lifetimes are orders of magnitude longer than surface plasmons and thus result in substantially lower absorption losses[17]. Earlier works explored hyperbolic modes in hBN via either nano-structuring[5, 10] or scattering-type scanning nearfield optical microscopy (s-SNOM)[6-8, 18-20], both of which are technically challenging experiments making characterizing the properties of polaritons in different crystals of 2D materials difficult. For example, it becomes arduous to assess the quality and dispersion of hBN crystals produced with different isotopic purities[19, 21], or fabricated using different growth techniques. It has historically also been difficult to measure hyperbolic polaritons in the lower Reststrahlen band of hBN due to the lack of commercial laser sources for s-SNOM. Indeed, this is even more problematic in investigating phonon polaritons in other two dimensional materials[22] such as transition metal dichalcogenides (TMDs), due to the inherently low-energy scale of their optic phonons[23, 24].

Here we demonstrate how infrared attenuated total reflectance (ATR) spectroscopy using an ATR microscope objective, can efficiently launch and measure the properties of hyperbolic phonon polaritons (HPhPs) and anisotropic dielectric resonances in a Kretschmann-Raether[25] configuration. In this approach, total internal reflection (TIR) at the boundary between a dielectric and a polaritonic material launches the polariton (Fig 1a), with an in-plane momentum determined by the incident angle and the index of refraction of the prism. The use of an ATR microscope objective enables measurements over a broad spectral range from small regions of interest, overcoming earlier limitations with prisms that are orders of magnitude larger than the material flake being studied. This ATR micro-spectroscopy approach enables broadband, quick measurements using an appropriate prism and FTIR/microscope configuration. For example, polaritons in any 2D material could be measured within the transmission window of germanium with the same prism and detectors used in this paper, including graphene[26] and transition metal oxides[23], while modifying the setup to use a DLaTGS or bolometer detectors would offer these measurements from TMDs[24].

In ATR spectroscopy a high refractive index prism provides a high in-plane momentum to couple into polariton modes, with wavevector $k=k_0n_1 sin(\theta')$ (illustrated in Fig. 1a). The polariton dispersion (frequency vs wavevector) of hyperbolic polaritons in an 800nm thick flake of hBN exfoliated onto a 3μm layer of $SiO_2$ on silicon is shown in Fig. 1b, calculated using a transfer-matrix-based approach[7]. The bright lines (corresponding to maxima in the imaginary part of the Fresnel coefficient) represent hyperbolic polariton modes. Note, this hyperbolic dispersion exhibits multiple branches, each with successively higher spatial confinement (wavelength compression) at the same frequency. This is a hallmark of hyperbolic media, with the number of branches and the corresponding slopes being strongly thickness dependent[7]. This contrasts with surface polaritons, which typically exhibit a single branch (e.g. one wavevector per incident frequency). Onto this HPhP dispersion curve we plot the dispersion of light in vacuum (cyan), the $SiO_2$ substrate (red) and the Ge prism (green solid line). We can couple into hyperbolic polaritons using a germanium prism (aka $k=k_0n_1$) in the region between the red and green lines by changing the angle of incidence. Crucially, we can potentially couple into a series of different hyperbolic modes, even at the incidence angles of typical Cassegrain microscope objectives (green dashed line Fig. 1b). At angles commensurate with those achieved in our experiment, we expect to see four resolvable polariton resonances in the upper Reststrahlen band, shown in Fig. 1c (calculated using a transfer matrix method (TMM)[27]), and three in the lower Reststrahlen. These are due to stimulation of the HPhPs within the first four (three) branches in the upper (lower) Reststrahlen regions. This provides strong motivation that this proposed approach can be used to systematically investigate polaritons in small flakes of 2D materials[22] using ATR micro-spectroscopy.

Our experimental approach to launching hyperbolic polaritons is based around a commercially available (Bruker Optics) ATR microscope objective, schematically shown in Fig 1a. A germanium prism ($n_1 \sim 4$) with a base diameter of ~100um is mounted at the focal point of a 20x Cassegrain-type reflective microscope objective (NA=0.6). This prism is pressed onto the surface of the sample with a force 0.5 N, ensuring good optical contact between the sample and the prism. Any non-uniformities in this interface will lead to a small gap between sample and prism, however these will be extremely small compared with the incident wavelength and thus can be neglected. The as-manufactured ATR objective provides light at incident angles between approximately 22 and 37°, thus potentially broadening the range of polaritonic wavevectors that are simultaneously launched and collected. This would potentially produce significant angular broadening in the collected spectra. In order to reduce the effects of angular spreading, we added a custom-made aperture plate to the objective, which restricted the incident angle to $\theta=33\pm4°$ (as labelled in Fig 1a)[28]. When infrared light enters the prism it refracts, slightly altering the angle incident on the sample (see inset to Fig 1a). To find the true incident angle at the prism-sample interface, we fit to our experimental data, finding a refraction corrected incident angle ($\theta'$) of approximately 39°. This aperture plate also allowed us to control the incident polarization to *s*- or *p*-polarized light as previously demonstrated[28]. Prior work has exploited prism-based techniques to measure the properties of hyperbolic materials,[29, 30] however, due to the low-refractive-index of the prisms used in those experiments they were unable to measure defined polariton resonances. Furthermore, they typically used samples and prisms on the scale of cm's – which are not suitable for the relatively small flakes typical of exfoliated hBN or other 2D materials. Thus, the requirement of achieving TIR places stringent requirements on the possible substrates for the polariton film and must be taken into account for any experimental design.

To address the choice of an appropriate substrate for the measurement, we consider the critical angle for TIR. The critical angle ($\theta_c$) is defined in terms of the refractive index of the prism ($n_1$) and of the substrate as ($n_2$)[31]:

$$\theta_c = \arcsin\left(\frac{n_2}{n_1}\right). \tag{1}$$

Our samples consist of flakes of hBN, mechanically exfoliated onto a thick $SiO_2$ layer grown on silicon. The critical angle for $SiO_2$ ($n\sim1.62$) is 24°, and Si ($n\sim3.4$) is 58°, which means that ATR measurements can be performed on $SiO_2$ and other low-index dielectric substrates, but that high index substrates such as silicon require incident angles that cannot be accessed within the current objective design. This has significant implications for sample preparation. For instance, while we exfoliated hBN layers onto thin films of $SiO_2$ grown on silicon, it is also important to ensure the evanescent wave launched at the prism-$SiO_2$ interface does not significantly interact with the silicon substrate. This is dictated by the decay length ($l$) of an evanescent field resulting from the TIR of the incident light with an incident free-space wavelength λ that is launched at the prism boundary and is given by the equation[31]:

$$l = \frac{\lambda}{2\pi\sqrt{(n_1\sin(\theta))^2 - n_2^2}} \quad (2)$$

For the case of $\theta=30$, $n_1 =$~4, $n_2 \sim 1.62$ and $\lambda\sim7\mu m$, the decay length is 950nm, so to ensure that no significant interactions with the silicon substrate occur, the oxide thickness must be a few times larger than this decay length. Thus, in our experiments we choose a SiO$_2$ thickness of 3 μm for our sample.

Three types of hBN crystals are studied in this work with natural isotopic concentration (~80% $^{11}$B/ ~20% $^{10}$B)[32], and isotopically enriched[19] to ~99% $^{10}$B and $^{11}$B. Details concerning the growth of the original bulk crystals from which the flakes were exfoliated are available in the literature[33]. Prior to IR characterization, the thickness of each flake was measured using atomic force microscopy (AFM), thereby enabling accurate comparison with electromagnetic simulations. As mentioned previously, this is especially pertinent for hyperbolic modes, as the volume-confined nature gives rise to a strong thickness dependence of the polariton dispersion[7]. Experiments were carried out using an IR microscope (Bruker Hyperion 2000) coupled to a Fourier-transform IR (FTIR) spectrometer (Bruker Vertex 70V), with a liquid-nitrogen-cooled HgCdTe (MCT) detector (Infrared Associates FTIR 24-0.25) and a SiC-glowbar source. To perform ATR measurements the sample was slowly brought into contact with the prism (0.5N, ~1mm/s) that is mounted to the ATR microscope objective and carefully lowered after measurement (~ 1mm/s), to minimize damage to the flake. After each measurement the prism was cleaned with isopropyl alcohol on a lens tissue. The spectra were collected using both unpolarized and polarized IR light via a germanium wire grid polarizer (Pike Technologies 090-1500) and referenced to the prism in contact with a gold mirror (to minimize attenuation from atmospheric water).

A representative, *p*-polarized ATR reflectance spectra of a 490nm thick flake of naturally abundant hBN is provided in Fig. 2a. We measure a series of resonant absorbing modes, in both the lower and upper Reststrahlen bands, as well as additional peaks that can be assigned to absorption within the silicon dioxide layer. These results compare extremely well with numerical simulations using both transfer matrix and finite element methods, suggesting that we can achieve good optical contact between prism and sample and high degree of angular selectivity. The small deviations between experiment and theory can be attributed to a combination of angular dispersion in the incident light and thickness non-uniformities of the hBN sample. To confirm that the resonant modes shown in Fig. 2a are hyperbolic modes in the hBN flake, we used FEM simulations to plot the electromagnetic field profiles (z- component of electric field) for each mode, shown in Fig 2c (i-vii). In the lower Reststrahlen band we observe two modes (i and ii), with different propagation angles and negative phase velocity[34], characteristic of Type 1 HPhPs in hBN. We note that the Type 1 Reststrahlen region overlaps with an absorption band in the underlying silicon dioxide, which enhances absorption in these modes. In the upper Reststrahlen band we observe three modes, again each with different propagation angles but instead with positive phase velocity, corresponding to Type 2 HPhPs. This demonstrates that ATR measurements can be used to directly measure hyperbolic polariton modes and thus provide access to the dispersion relationship.

In addition to the hyperbolic modes, in s-polarized reflectance (Fig. 2b) we observe two peaks close to the upper Reststrahlen band of hBN. In this spectral region all components of the hBN dielectric function are positive, so they cannot be associated with polaritons. However, below the TO phonon energy the dielectric constant of a phononic material can become extremely large and positive, with a refractive index much larger than that in non-dispersive materials[35]. This means that dielectric cavity, Mie resonances[36] can be formed in extremely sub-wavelength cavities that can be comparable in size to polaritonic structures[37]. For clarity we note that Mie resonances, despite the nanoscale size of the structures cannot be considered sub-diffractional, as the wavelength compression is due purely to the extremely high index of refraction near this TO phonon absorption band. In our case we observe both first and second order dielectric resonances (Fig 2b,c, vi and vii), due to the refractive index of approximately 7.9 and 15.8 at these two resonant frequencies, respectively, which as stated is significantly higher than can be obtained from typical, weakly dispersing dielectric materials. Such highly anisotropic Mie resonances have not been measured previously, and could potentially be used for the creation of nanoscale dielectric Mie-type resonators in the IR[38, 39], for instance enabling Huygen's mode based devices[40] or perfect absorbing coatings[41].

As the angle of the incident light is difficult to vary in our microscope-based setup, we cannot systematically plot out the dispersion curve for a given flake as the in-plane momentum cannot be controlled. We note that with more advanced objective designs that a variable incidence angle could potentially be realized. Instead, we can study the dispersion of each mode using different thicknesses of boron nitride. Unlike in conventional polaritonic systems, where the mode frequency is relatively insensitive to film thickness, as mentioned above, hyperbolic modes are extremely sensitive to such changes. This is because each HPhP in the spectrum can be considered an 'etalon-like' mode since hyperbolic polaritons propagate within the bulk material at a distinct, frequency-dependent propagation angle[3, 42, 43]. However, the etalon modes are dictated by the highly compressed HPhP wavelength rather than the free-space value[7]. Therefore, by changing the thickness of the hBN flake, we can also tune the modal spectrum (Fig. 3d-f), as has been shown in earlier work[7]. To demonstrate this, we measure a series of different thicknesses of hBN flakes in the Type I HPhP (p-pol, Fig. 3a), dielectric (s-pol, Fig. 3b) and II HPhP regimes (p-pol Fig. 3c). We do not include thickness varying data for s-pol in the Type I region as only absorption from the SiO$_2$ substrate was observed. All the modes discussed above were observed to continuously tune as the thickness is increased, with the thickness of the film also determining the number of modes observable in experiments. This is consistent with an etalon-like behaviour of both hyperbolic and dielectric modes in the hBN films. Because of this, it is also possible to use this spectral response to extract the corresponding dielectric function of these materials using variable thickness ATR spectroscopy. We subsequently fit the peak positions in Fig. 3a-c, and compare the peak positions to a numerical model in Fig. 3d-f, showing that the trends observed in experiments provide good quantitative agreement with our calculated results. While the polaritonic modes in hBN are extremely sensitive to both thickness and incident angle, the dielectric modes are weakly dependant on incidence angle. This provides a useful approach to calibrating the correct incidence angle to use for a given ATR objective prism.

To illustrate the potential of this technique for distinguishing different materials, determining their hyperbolic properties and the potential for extracting the IR dielectric function, we compare hBN with three different isotopic purities of boron; 20% $^{10}$B, 80% $^{11}$B (naturally abundant), ~99% $^{10}$B and ~99% $^{11}$B. In Fig. 4a we show an ATR reflectance spectrum for three such flakes with similar thicknesses. Due to the shifting of the TO phonon energy with isotopic concentration[19], the resonances of the $^{11}$B- and $^{10}$B-enriched hBN, red and blue shift with respect to natural hBN, respectively. We also note a reduction in the resonance linewidth that is due to reduced phonon scattering in isotopically enriched materials. This result illustrates that the ATR approach developed in this paper can be used to launch and measure polaritons in different materials with different isotopic concentrations. In principle, by implementing a least-squares fitting program and considering several thicknesses of hBN flakes, it would be possible to extract the dielectric parameters of the material studied. However, we note from our transfer matrix calculations in Fig. 3d-e that when the thickness of the hBN films tend to zero, the resonance in the ATR spectra becomes located at the TO phonon frequency and strongly absorbing. The significant absorption in thin films of hBN is attributed to the high photonic density of states in a hyperbolic material. By fitting the calculated absorption line (shown in Fig. 4b inset) with a Lorentzian lineshape, we can extract the full width half maximum ($\Gamma$), and plot this as a function of film thickness (Fig. 4b). We observe that $\Gamma$ increases linearly with thickness, and by extrapolating down to zero thickness we can infer the damping constant of the hBN film. The damping constant can be approximated at zero thickness because a polariton absorption linewidth is a combination of both radiative and material losses[25] – and as the film thickness tends to zero, light cannot couple to the hBN – resulting in a linewidth that trends to the intrinsic material loss. This provides a relatively simple approach to estimating the damping constant in Type II hyperbolic thin films.

In conclusion, we have proposed and demonstrated the ability to implement ATR micro-spectroscopy as a means for launching and probing hyperbolic polaritons and dielectric resonances in 2D materials. This approach relies on both careful definition of the incident angle and polarization state. Using this method, we measured three distinct HPhPs in the upper, and two in the lower Reststrahlen band of hBN, as confirmed by comparisons with numerical models. Furthermore, we observed highly anisotropy, extreme-index dielectric resonances that have not been detected previously. Subsequently, by measuring a range of flakes with different thicknesses, we were able to show the dispersing nature of hBN polaritons and that our technique can distinguish flakes of different materials, demonstrated here for varying isotopic purity hBN. Finally, we used electromagnetic simulations to show that this approach could be used to estimate the dielectric properties of hBN flakes. Whilst this technique has limitations, most notably the single incident angle, the broad spectral range allows flexible, simple investigations of a range of different materials of varying size, shape or polariton frequencies. This makes this technique the only means of providing quick, high throughput measurements of a range of different potential hyperbolic materials.

1. S.A. Maier: Plasmonics: Fundamentals and Applications, (Berlin, 2007).
2. D.N. Basov, M.M. Fogler and F.J. García de Abajo: Polaritons in van der Waals materials. *Science* **354**, 195 (2016).
3. A. Poddubny, I. Iorsh, P. Belov and Y. Kivshar: Hyperbolic metamaterials. *Nature Photonics* **7**, 948 (2013).
4. M. Noginov, M. Lapine, V.A. Podolskiy and Y. Kivshar: Focus issue: hyperbolic metamaterials. *Optics Express* **21**, 14895 (2013).
5. J.D. Caldwell, A. Kretinin, Y. Chen, V. Giannini, M.M. Fogler, Y. Francescato, C. Ellis, J.G. Tischler, C. Woods, A.J. Giles, M. Hong, K. Watanabe, T. Taniguchi, S.A. Maier and K.S. Novoselov: Sub-diffractional, Volume-confined Polaritons in the Natural Hyperbolic Material Hexagonal Boron Nitride. *Nature Communications* **5**, 5221 (2014).
6. P. Li, M. Lewin, A.V. Kretinin, J.D. Caldwell, K.S. Novoselov, T. Taniguchi, K. Watanabe, F. Gaussmann and T. Taubner: Hyperbolic Phonon-Polaritons in Boron Nitride for near-field optical imaging and focusing. *Nature Communications* **6**, 7507 (2015).
7. S. Dai, Z. Fei, Q. Ma, A.S. Rodin, M. Wagner, A.S. McLeod, M.K. Liu, W. Gannett, W. Regan, M. Thiemens, G. Dominguez, A.H. Castro Neto, A. Zettl, F. Keilmann, P. Jarillo-Herrero, M.M. Fogler and D.N. Basov: Tunable phonon polaritons in atomically thin van der Waals crystals of boron nitride. *Science (Washington)* **343**, 1125 (2014).
8. S. Dai, Q. Ma, T. Anderson, A.S. McLeod, Z. Fei, M.K. Liu, M. Wagner, K. Watanabe, T. Taniguchi, M. Thiemens, F. Keilmann, P. Jarillo-Herrero, M.M. Fogler and D.N. Basov: Subdiffractional focusing and guiding of polaritonic rays in a natural hyperbolic material. *Nature Communications* **6**, 6963 (2015).
9. Z. Liu, H. Lee, Y. Xiong, C. Sun and X. Zhang: Far-field optical hyperlens magnifying sub-diffraction limited objects. *Science* **315**, 1686 (2007).
10. F.J. Alfaro-Mozaz, P. Alonso-González, S. Vélez, I. Dolado, M. Autore, S. Mastel, F. Casanova, L.E. Hueso, P. Li, A.Y. Nikitin and R. Hillenbrand: Nanoimaging of resonating hyperbolic polaritons in linear boron nitride antennas. *Nature Communications* **8**, 15624 (2017).
11. T.G. Folland, A. Fali, S.T. White, J.R. Matson, S. Liu, N.A. Aghamiri, J.H. Edgar, R.F. Haglund, Y. Abate and J.D. Caldwell: Reconfigurable Mid-Infrared Hyperbolic Metasurfaces using Phase-Change Materials, (arXiv:1805.08292, 2018).
12. M. Autore, P. Li, I. Dolado, F.J. Alfaro-Mozaz, R. Esteban, A. Atxabal, F. Casanova, L.E. Hueso, P. Alonso-González, J. Aizpurua, A.Y. Nikitin, S. Vélez and R. Hillenbrand: Boron nitride nanoresonators for phonon-enhanced molecular vibrational spectroscopy at the strong coupling limit. *Light: Science &Amp; Applications* **7**, 17172 (2018).
13. K.V. Sreekanth, Y. Alapan, M. ElKabbash, E. Ilker, M. Hinczewski, U.A. Gurkan, A. De Luca and G. Strangi: Extreme sensitivity biosensing platform based on hyperbolic metamaterials. *Nature Materials* **15**, 621 (2016).
14. A.J. Hoffman, L. Alekseyev, S.S. Howard, K.J. Franz, D. Wasserman, V.A. Podolskiy, E.E. Narimanov, D.L. Sivco and C. Gmachl: Negative Refraction in semiconductor metamaterials. *Nature Materials* **6**, 946 (2007).
15. J. Yao, Z. Liu, Y. Liu, Y. Wang, C. Sun, G. Bartal, A.M. Stacy and X. Zhang: Optical Negative Refraction in Bulk Metamaterials of Nanowires. *Science* **321**, 930 (2008).
16. K. Korzeb, M. Gajc and D.A. Pawlak: Compendium of natural hyperbolic materials. *Optics Express* **23**, 25406 (2015).
17. J.D. Caldwell, L. Lindsey, V. Giannini, I. Vurgaftman, T. Reinecke, S.A. Maier and O.J. Glembocki: Low-Loss, Infrared and Terahertz Nanophotonics with Surface Phonon Polaritons. *Nanophotonics* **4**, 44 (2015).
18. L.V. Brown, M. Davanco, Z. Sun, A. Kretinin, Y. Chen, J.R. Matson, I. Vurgaftman, N. Sharac, A.J. Giles, M.M. Fogler, T. Taniguchi, K. Watanabe, K.S. Novoselov, S.A. Maier, A. Centrone and J.D. Caldwell: Nanoscale Mapping and Spectroscopy of Nonradiative Hyperbolic Modes in Hexagonal Boron Nitride Nanostructures. *Nano Letters* **18**, 1628 (2018).

19. A.J. Giles, S. Dai, I. Vurgaftman, T. Hoffman, S. Liu, L. Lindsay, C.T. Ellis, N. Assefa, I. Chatzakis, T.L. Reinecke, J.G. Tischler, M.M. Fogler, J.H. Edgar, D.N. Basov and J.D. Caldwell: Ultralow-loss polaritons in isotopically pure boron nitride. *Nature Materials* **17**, 134 (2018).
20. P. Li, I. Dolado, F.J. Alfaro-Mozaz, F. Casanova, L.E. Hueso, S. Liu, J.H. Edgar, A.Y. Nikitin, S. Vélez and R. Hillenbrand: Infrared hyperbolic metasurface based on nanostructured van der Waals materials. *Science* **359**, 892 (2018).
21. T. Vuong, S. Liu, A. Van der Lee, R. Cuscó, L. Artús, T. Michel, P. Valvin, J. Edgar, G. Cassabois and B. Gil: Isotope engineering of van der Waals interactions in hexagonal boron nitride. *Nature materials* **17**, 152 (2018).
22. T. Low, A. Chaves, J.D. Caldwell, A. Kumar, N.X. Fang, P. Avouris, T.F. Heinz, F. Guinea, L. Martin-Moreno and F.H.L. Koppens: Polaritons in layered two-dimensional materials. *Nature Materials* **16**, 182 (2017).
23. Z. Zebo, C. Jianing, W. Yu, W. Ximiao, C. Xiaobo, L. Pengyi, X. Jianbin, X. Weiguang, C. Huanjun, D. Shaozhi and X. Ningsheng: Highly Confined and Tunable Hyperbolic Phonon Polaritons in Van Der Waals Semiconducting Transition Metal Oxides. *Advanced Materials* **30**, 1705318 (2018).
24. S.-i. Uchida and S. Tanaka: Optical Phonon Modes and Localized Effective Charges of Transition-Metal Dichalcogenides. *Journal of the Physical Society of Japan* **45**, 153 (1978).
25. H. Raether: Surface plasmons on smooth and rough surfaces and on gratings, (Springer-Verlag, Berlin ; New York, 1988).
26. X. Dai, L. Jiang and Y. Xiang: Tunable THz Angular/Frequency Filters in the Modified Kretschmann Raether Configuration With the Insertion of Single Layer Graphene. *IEEE Photonics Journal* **7**, 1 (2015).
27. N.C. Passler and A. Paarmann: Generalized 4 × 4 matrix formalism for light propagation in anisotropic stratified media: study of surface phonon polaritons in polar dielectric heterostructures. *Journal of the Optical Society of America B* **34**, 2128 (2017).
28. T.W.W. Maß and T. Taubner: Incident Angle-Tuning of Infrared Antenna Array Resonances for Molecular Sensing. *ACS Photonics* **2**, 1498 (2015).
29. L. Luo and T. Tang: Goos-Hänchen effect in Kretschmann configuration with hyperbolic metamaterials. *Superlattices and Microstructures* **94**, 85 (2016).
30. C. Zhang, N. Hong, C. Ji, W. Zhu, X. Chen, A. Agrawal, Z. Zhang, T.E. Tiwald, S. Schoeche, J.N. Hilfiker, L.J. Guo and H.J. Lezec: Robust Extraction of Hyperbolic Metamaterial Permittivity Using Total Internal Reflection Ellipsometry. *ACS Photonics* **5**, 2234 (2018).
31. M. Born and E. Wolf: Principles of optics : electromagnetic theory of propagation, interference and diffraction of light. 808 (1980).
32. T. Taniguchi and K. Watanabe: Synthesis of high-purity boron nitride single crystals under high pressure by using Ba-BN solvent. *Journal of Crystal Growth* **303**, 525 (2007).
33. S. Liu, R. He, L. Xu, J. Li, B. Liu and J.H. Edgar: Large-scale growth of monoisotopic hexagonal boron nitride single crystals, (submitted, 2018).
34. E. Yoxall, M. Schnell, A.Y. Nikitin, O. Txoperena, A. Woessner, M.B. Lundeberg, F. Casanova, L.E. Hueso, F.H.L. Koppens and R. Hillenbrand: Direct observation of ultraslow hyperbolic polariton propagation with negative phase velocity. *Nature Photonics* **9**, 674 (2015).
35. J.A. Schuller, R. Zia, T. Taubner and M.L. Brongersma: Dielectric Metamaterials Based on Electric and Magnetic Resonances of Silicon Carbide Particles. *Physical Review Letters* **99**, 107401 (2007).
36. A.I. Kuznetsov, A.E. Miroshnichenko, M.L. Brongersma, Y.S. Kivshar and B. Luk'yanchuk: Optically resonant dielectric nanostructures. *Science* **354** (2016).
37. J. Caldwell and O. Glembocki: Low-loss, extreme subdiffraction photon confinement via silicon carbide localized surface phonon polariton resonators. *Nano …* (2013).
38. I. Staude and J. Schilling: Metamaterial-inspired silicon nanophotonics. *Nature Photonics* **11**, 274 (2017).

39. J.C. Ginn, I. Brener, D.W. Peters, J.R. Wendt, J.O. Stevens, P.F. Hines, L.I. Basilio, L.K. Warne, J.F. Ihlefeld, P.G. Clem and M.B. Sinclair: Realizing Optical Magnetism from Dielectric Metamaterials. *Physical Review Letters* **108**, 097402 (2012).
40. A. Howes, W. Wang, I. Kravchenko and J. Valentine: Dynamic transmission control based on all-dielectric Huygens metasurfaces. *Optica* **5**, 787 (2018).
41. W. Li and J. Valentine: Metamaterial Perfect Absorber Based Hot Electron Photodetection. *Nano Letters* **14**, 3510 (2014).
42. A.J. Giles, S. Dai, O.J. Glembocki, A.V. Kretinin, Z. Sun, C.T. Ellis, J.G. Tischler, T. Taniguchi, K. Watanabe, M.M. Fogler, K.S. Novoselov, D.N. Basov and J.D. Caldwell: Imaging of Anomalous Internal Reflections of Hyperbolic Phonon-Polaritons in Hexagonal Boron Nitride. *Nano Letters* **16**, 3858 (2016).
43. S. Ishii, A.V. Kildishev, E.E. Narimanov, V.M. Shalaev and V.P. Drachev: Sub-wavelength interference pattern from volume plasmon polaritons in a hyperbolic medium. *Laser and Photonics Reviews* **7**, 265 (2013).

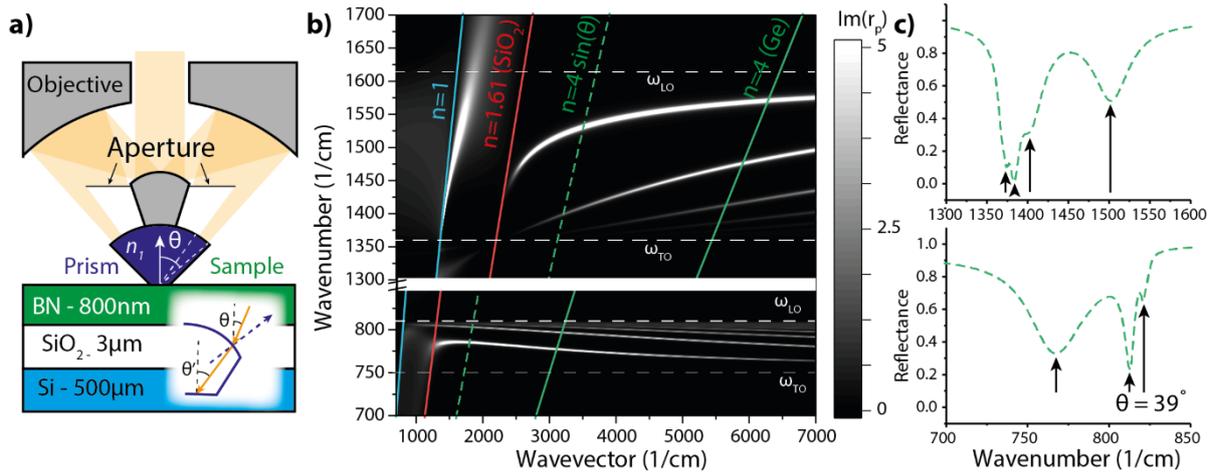

*Figure 1: a) A schematic of the prism-coupled, ATR micro-spectroscopy measurements of hyperbolic polaritons in hBN. The schematic details the basic design principle of the ATR objective used to measure small (approx. 100um across) flakes of hBN for this study. Inset: Illustration of how refraction at the non-planar prism-air interface can result in a slightly different incidence angle in the prism vs in free space. b) The dispersion of HPhPs in hBN are represented in a contour plot with the z-axis plotting the imaginary part of the Fresnel coefficient. For assistance with the discussion in the text, the dispersion of light in vacuum, SiO$_2$ and Germanium are provided as the blue, red and green solid lines, respectively. The green dashed line provides the dispersion of light through a typical Cassegrain-type objective using a Ge prism. (c) Provides the ATR reflectance spectrum calculated using transfer matrix methods for an incidence angle commensurate with experiments in both lower (top plot) and upper (bottom plot) Reststrahlen bands. Arrows indicate polariton modes.*

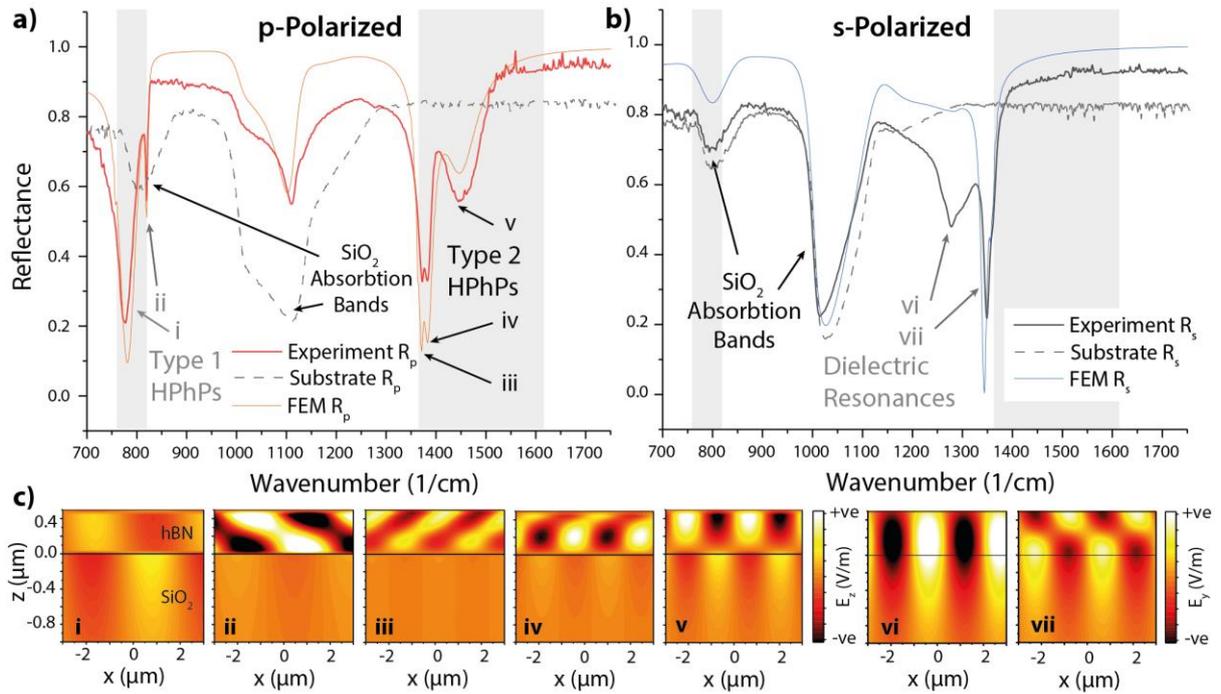

*Figure 2: a) p- (red solid curve) and b) s-polarized (grey solid curve) ATR micro-spectroscopy measurement of a 490nm thick hBN flake. Corresponding numerical calculations for this sample under both polarizations are provided as the orange and blue solid lines in a) and b), respectively. Both finite element (FEM) and transfer matrix techniques were performed, but as no appreciable differences were observed between the two calculated reflection spectra, only the former is plotted here. The grey shaded areas indicate the regions of the lower and upper Reststrahlen bands of hBN, where this material is naturally hyperbolic. Multiple resonances are observed, which can be identified as Type 1, Type 2 and dielectric resonances [labelled in a) and b)] by examining electromagnetic field profiles provided in c), with each mode designated in both a,b and c by the roman numerals i-v.*

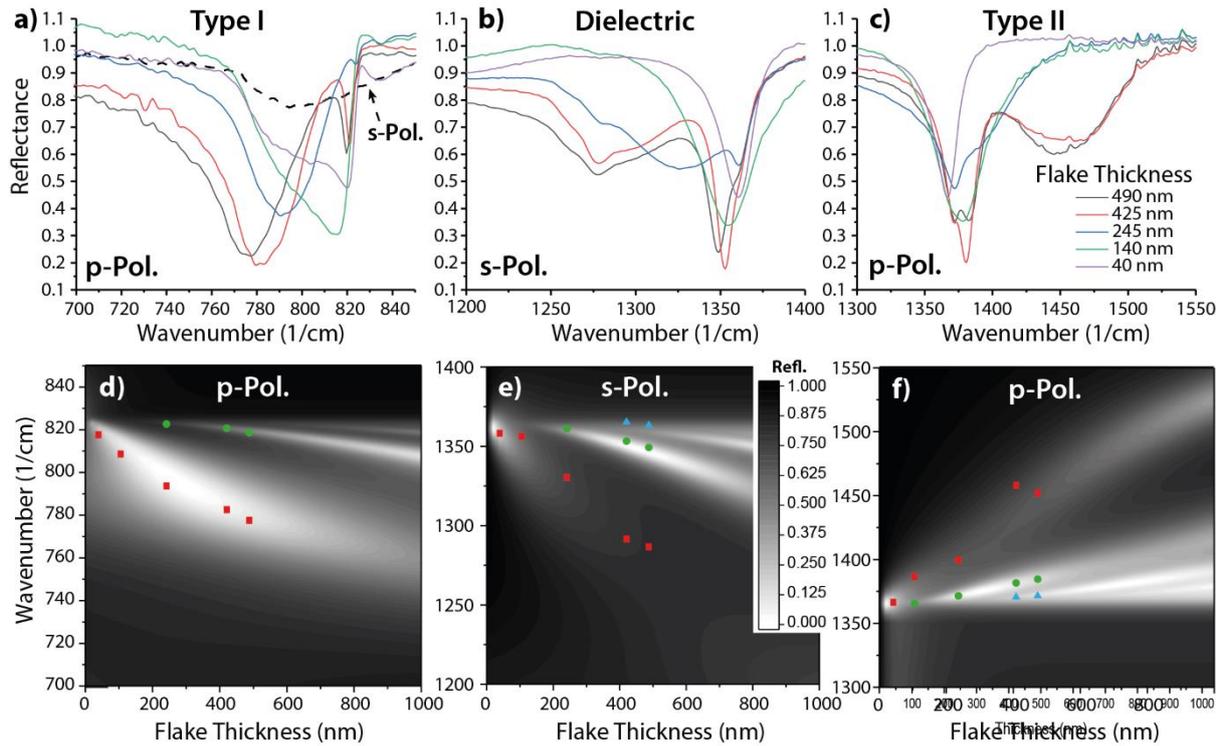

*Figure 3: Thickness dependence of ATR modes. a) – c) show experimentally measure reflectance spectra of various thicknesses of natural hBN flakes within the a) lower Reststrahlen (p-polarized), anisotropic dielectric (s-polarized) and c) upper Reststrahlen (p-polarized) bands. Contour plots of the simulated reflectance spectra for a range of hBN thicknesses in the are provided in d)-f) for the same spectral regions as in a)-c), respectively. As the thickness increases, the modes tune in frequency and the number of resolvable modes increases, mirroring the dispersion of Fig. 1b. Symbols in d)-f) correspond to the position of the peaks extracted from the experimental data presented in a)-c). Red Squares represent 1st order modes, Green circles 2nd order, and cyan triangles 3rd order.*

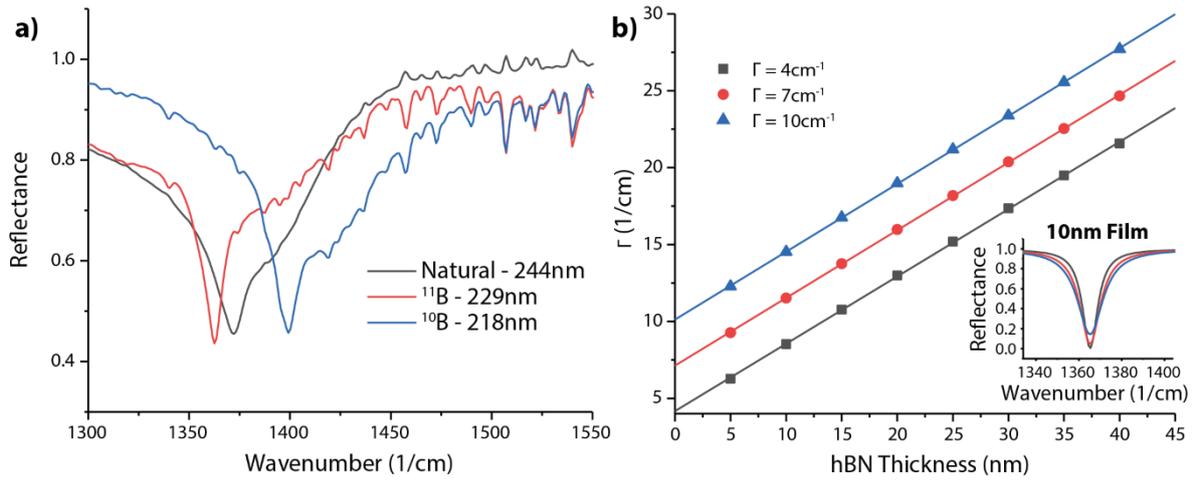

*Figure 4: Identification and measurement of isotopically enriched materials using ATR micro-spectroscopy. a) Measurements of three flakes of hBN with different isotopic purities, showing the spectral shift in the phonon polariton associated with a change in the TO phonon frequency. Corresponding thicknesses are provided in the legend. b) Dependence of the full width half maximum (Γ) of the reflection dip in the ATR spectra presented in a), demonstrating that as the film tends to zero thickness that the width of the resonance is approximately equal to the damping of the hBN.*